\documentclass[journal=jacsat,manuscript=article]{achemso}
\usepackage{epstopdf}
\usepackage{amsmath}
\usepackage[version=3]{mhchem} 

\author{O. Del Pozo-Zamudio}
\email{o.delpozo@sheffield.ac.uk}
\affiliation[University of Sheffield]{Department of Physics and Astronomy, University of Sheffield, Sheffield S3 7RH, United Kingdom}
\author{S. Schwarz}
\affiliation[University of Sheffield]{Department of Physics and Astronomy, University of Sheffield, Sheffield S3 7RH, United Kingdom}
\author{M. Sich}
\affiliation[University of Sheffield]{Department of Physics and Astronomy, University of Sheffield, Sheffield S3 7RH, United Kingdom}
\author{I. A. Akimov}
\affiliation[Technische Universit\"at Dortmund]{Experimentelle Physik 2, Technische Universit\"at Dortmund, 44227 Dortmund, Germany}
\author{M. Bayer}
\affiliation[Technische Universit\"at Dortmund]{Experimentelle Physik 2, Technische Universit\"at Dortmund, 44227 Dortmund, Germany}
\author{R. C. Schofield}
\affiliation[University of Sheffield]{Department of Physics and Astronomy, University of Sheffield, Sheffield S3 7RH, United Kingdom}
\author{E. A. Chekhovich}
\affiliation[University of Sheffield]{Department of Physics and Astronomy, University of Sheffield, Sheffield S3 7RH, United Kingdom}
\author{B. J. Robinson}
\affiliation[University of Lancaster]{Department of Physics, University of Lancaster, Lancaster LA1 4YB, United Kingdom}
\author{N. D. Kay}
\affiliation[University of Lancaster]{Department of Physics, University of Lancaster, Lancaster LA1 4YB, United Kingdom}
\author{O. V. Kolosov}
\affiliation[University of Lancaster]{Department of Physics, University of Lancaster, Lancaster LA1 4YB, United Kingdom}
\author{A. I. Dmitriev }
\affiliation[I. M. Frantsevich Institute for Problems of Material Science]{I. M. Frantsevich Institute for Problems of Material Science, NASU, Kiev-142, Ukraine}
\author{G. V. Lashkarev}
\affiliation[I. M. Frantsevich Institute for Problems of Material Science]{I. M. Frantsevich Institute for Problems of Material Science, NASU, Kiev-142, Ukraine}
\author{D. N. Borisenko}
\affiliation[Institute of Solid State Physics]{Institute of Solid State Physics, Russian Academy of Sciences, Chernogolovka 142432, Russia}
\author{N. N. Kolesnikov}
\affiliation[Institute of Solid State Physics]{Institute of Solid State Physics, Russian Academy of Sciences, Chernogolovka 142432, Russia}
\author{A. I. Tartakovskii}
\email{a.tartakovskii@sheffield.ac.uk}
\affiliation[University of Sheffield]{Department of Physics and Astronomy, University of Sheffield, Sheffield S3 7RH, United Kingdom}

\title{Optical properties of two-dimensional gallium chalcogenide films}

\begin{document}


\begin{abstract}
Gallium chalcogenides are promising building blocks for novel van der Waals heterostructures. We report on low-temperature micro-photoluminescence (PL) of GaTe and GaSe films with thicknesses ranging from 200 nm to a single unit cell. In both materials, PL shows dramatic decrease by 10$^4$-10$^5$ when film thickness is reduced from 200 to 10 nm. Based on evidence from cw and time-resolved PL, we propose a model explaining the PL decrease as a result of non-radiative carrier escape via surface states. Our results emphasize the need for special passivation of two-dimensional films for optoelectronic applications.
\end{abstract}

\section{Introduction}

The discovery and research of remarkable properties of two-dimensional (2D) sheets of carbon \cite{NovoselovScience2004}, known as graphene, has sparked interest in other layered materials such as metal chalcogenides (MCs) \cite{NovoselovPNAS2005,WangNatNano2012}. Recent progress in device fabrication opens up new possibilities for 2D MC films in nano- and opto-electronics \cite{WangNatNano2012,BritnellScience2013,GeimNature2013,RadisavljevicNatNano2011,WuArxiv2015,WithersNatMater2015}. Recent efforts in research of 2D semiconductors and their devices focused mainly on molybdenum and tungsten dichalcogenides\cite{MakPRL2010,SplendianiNanoLett2010,EdaNanoLett2011,BritnellScience2013,JonesNatNano2013,RossNatComm2013,WithersNatMater2015} and relatively little has been reported on optical properties of thin films of III-VI materials: only recently electronic\cite{HuACSNano2012,LateAdvMater2012,BritnellScience2013,LiuACSNano2014} and photonic\cite{SchwarzNanoLett2014} trial devices have been reported for GaSe and GaTe\cite{LiuACSNano2014}, and room-temperature photoluminescence (PL) was measured in detail for InSe thin films \cite{MuddAdvMat2013}. As the crystal growth and improvement of thin GaSe films continues \cite{HuACSNano2012,LeiNanoLett2013,KokhCrysResTech2011}, further exploration of physical properties of III-VI materials presented in this work is motivated by their potential use as building blocks for novel van der Waals heterostructures, where materials with a range of bandgaps and band offsets will be required \cite{GeimNature2013,WithersNatMater2015}. Furthermore, in contrast to molybdenum and tungsten chalcogenides emitting light efficiently only in films with a single unit cell thickness, III-VI materials are bright light emitters in a range of thicknesses\cite{MuddAdvMat2013}. This may relax the stringent fabrication requirements and add flexibility for the novel heterostructured devices such as light emitting diodes \cite{WithersNatMater2015}. 

Both GaTe and GaSe are layered crystals with strong covalent in-plane inter-atomic bonding (with some ionic contributions \cite{CamaraPRB2002,WanPRB1992}) and weaker predominantly van der Waals inter-plane bonding \cite{Hulliger1976,CapozziPRB1981,ZolyomiPRB2013,CamaraPRB2002,YamamotoPRB2001}. Single tetralayer having hexagonal in-plane structure consists of two Ga atoms and two Se or Te atoms: Se-Ga-Ga-Se and Te-Ga-Ga-Te \cite{ZolyomiPRB2013}. The bulk lattices are built by stacking tetralayers, which can occur in several ways\cite{Hulliger1976,CapozziPRB1981,CamaraPRB2002}. For the wider studied GaSe, several types of stacking exist leading to different polytypes\cite{Hulliger1976,CapozziPRB1981,CamaraPRB2002}. For GaTe having a monoclinic crystal lattice \cite{Grasso1975,WanPRB1992} the polytypic behavior has not been observed \cite{YamamotoPRB2001,WanPRB1992}. This may lead to lower probability of stacking faults in GaTe resulting in clearer observation (compared to GaSe) of excitonic features in optical experiments \cite{YamamotoPRB2001,WanPRB1992}.

Here we study optical properties of GaSe and GaTe thin films as a function of the film thickness. Continuous-wave (cw) and time-resolved low-temperature micro-PL for a wide range of films from 200 nm to one tetralayer thicknesses is measured. PL intensity is used to monitor the quantum yield (QY), which falls dramatically for thin films: integrated cw PL intensity drops by up to $\approx 10^4-10^5$ when the film thickness is reduced from 200 to 10 nm. A similar observation of reduced PL for thin films was previously reported for InSe and was explained as transition to a band-structure with an indirect bandgap, the conclusion also based on the observed PL blue-shift with the decreasing film thickness \cite{MuddAdvMat2013}. Such indirect bandgap behavior is also theoretically predicted for single monolayers of GaSe and GaTe \cite{ZolyomiPRB2013}. However, no size-quantization effects as in InSe are observed in our work for GaSe and GaTe. Based on the evidence from both cw and time-resolved spectroscopy, we develop a model that shows that the observed PL reduction can be explained by non-radiative carrier escape to surface states. Our explanation does not require introduction of the direct-to-indirect band-gap transition. Following the cw PL data analysis, we identify a critical film thickness of about 30-40 nm, below which the non-radiative carrier escape changes its character. The importance of surface states predicted by our results emphasizes the need for development of novel surface passivation for III-VI films, possibly involving oxygen-free dielectrics such as boron nitride \cite{KretininNanoLett2014}. 

\section{Experimental procedure}

\subsection{Fabrication of GaTe and GaSe samples}
Single crystals of GaSe and GaTe were grown by high-pressure vertical zone melting in graphite crucibles under Ar pressure. The detailed description of crystal growth processes and properties of GaSe and GaTe can be found in Refs.\cite{Kolesnikov2007,Kolesnikov2008,Borisenko2011,Kolesnikov2013}. The gallium mono-chalcogenides used in this work were synthesized from high-purity materials: Ga and Te - 99.9999 $\%$, Se - 99.9995 $\%$. The crystals used are high-resistivity semiconductors with low free carrier absorption, which has been confirmed in  infra-red transmition measurements. GaSe and GaTe have n-type and p-type conductivity, respectively. This is a typical observation: selenides are usually of n-type conductivity, whereas tellurides often can have conductivity of both types, even within one ingot. Such behavior is attributed to deviations of the crystal composition from stoichiometry, usual for metal chalcogenides. In our case the n-type conductivity clearly indicates some excess of Ga (donor) in GaSe, whereas the p-type conductivity indicates a slight excess of Te (acceptor) in the GaTe.

\begin{figure}[t]
\includegraphics[width=0.9\textwidth]{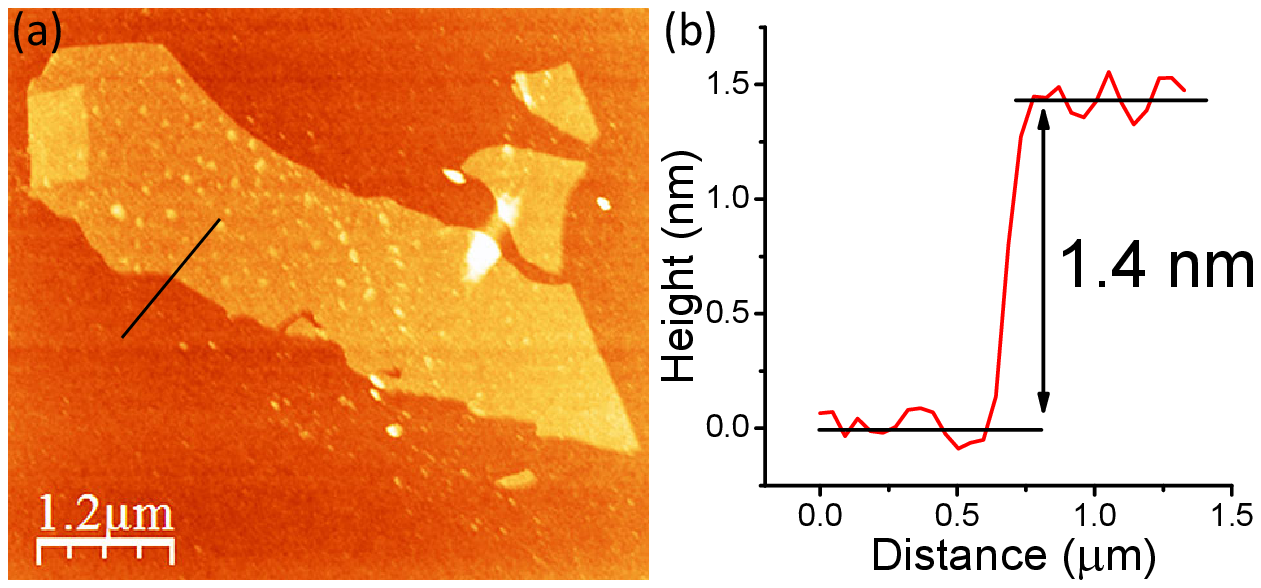}
   \caption{\label{Fig1} (a) AFM image of a GaSe thin film having a single unit cell thickness. This film is obtained by mechanical exfoliation. (b) Cross-sectional plot along the black line in (a). \\ \quad \\ \quad \\ \quad \\ \quad \\ \quad \\ \quad \\ \quad }

\end{figure}

The III-VI thin films studied in this work were fabricated by mechanical cleaving from bulk. The films were deposited on Si/SiO$_2$ substrates. Within the first 15 minutes after the exfoliation/deposition procedure, the  films were placed in a plasma-enhanced chemical vapor deposition (PECVD) reactor and a 15 nm Si$_3$N$_4$ layer was deposited with the sample maintained at a temperature of 300 $^\circ$C. This process leads to a complete encapsulation of the films, protecting them from interaction with oxygen and water present in the atmosphere. Films with a wide variety of thicknesses were obtained from single unit cell (single monolayer, ML) shown in Fig.\ref{Fig1} to 200 nm. The thicknesses of the films were determined using atomic force microscopy.

\subsection{Optical characterization methods}

Optical characterization of the GaTe and GaSe thin films was carried out using a low-temperature micro-photoluminescence ($\mu$PL) technique. The sample was placed on a cold finger in a continuous flow He cryostat at a temperature of 10 K. A microscope objective with NA of 0.6 was placed outside the cryostat and was used to focus the laser beam on the sample (in a $\approx$2$\mu$m spot) and to collect the photoluminescence (PL) from the films. In continuous-wave (cw) experiments, PL was detected with a 0.5~m spectrometer and a liquid nitrogen cooled charge coupled device.  For cw PL excitation a laser emitting at 532 nm (2.33 eV) was used. In the ultra-fast spectroscopy experiments the excitation of GaTe and GaSe layers was performed using frequency-doubled titanium-sapphire (wavelength of 415 nm, pulse duration $\approx 200$ ps) focused on the sample in $\approx$10$\mu$m spot. Time-resolved PL (TRPL) was detected using a streak camera. The temporal resolution of the experimental setup for the time-resolved measurements was 10 ps.

\section{Experimental results}

\subsection{Low-temperature cw PL results: GaTe}

Fig.\ref{Fig2}(a) shows typical PL spectra measured for Si$_3$N$_4$-capped GaTe thin films at $T=10$K (cw laser power $P$=2 mW in Fig.\ref{Fig2}(a)). In this figure, in all films but the one with the thickness $h_{film}$=8 nm, a narrow feature is observed around 1.75 eV. It is observed at an energy where free exciton (FE) PL is expected. We will therefore refer to such features in GaTe (and GaSe) films as a 'free exciton' peak as opposed to the low energy broad PL bands corresponding to excitons bound to impurities/defects and observed in the range of 1.6-1.7 eV for GaTe. The 'free excitons' may also experience disordered potential and degree of localization as evidenced from a relatively broad FE PL line of 10-15 meV (varying from sample-to-sample). A different behavior of the FE peak compared to the bound excitons has been verified in temperature and power-dependent measurements and is further confirmed in time-resolved studies discussed below. The FE peak is usually more pronounced in thick films of around 100 nm and above. In thin films as in the 8 nm film in the figure, the FE peak could only be observed under high power pulsed excitation, when the impurity/defect states saturate. The sharpest PL feature in GaTe spectra, the FE line, has the peak energy varying from film to film in the range 1.74-1.76 eV showing insignificance of size-quantization effects in contrast to InSe where the PL blue-shift was observed for thin layers \cite{MuddAdvMat2013}. 

A pronounced feature of PL measured from different films is a dramatic decrease of PL intensity with the decreasing thickness of the material [see Fig.\ref{Fig2}(b)]: about 10$^5$ (10$^4$) decrease is observed between 200 (100) and 7 nm. The strongest PL reduction by 3 orders of magnitude is detected between 200 and 40 nm. For $h_{film}<$40 nm, the PL intensity reduction slows down and decreases less than 100 times when $h_{film}$ is varied between 40 and 7 nm. The dotted curve in the graph shows the expected PL intensity behavior assuming thickness-independent quantum efficiency, i.e. when reduction in PL is caused solely by the reduced absorption and reduced number of e-h pairs created by the laser. The curve is described by the expression $I_{PL}=I_{GaTe} [1-exp(-\alpha_{GaTe} h_{film})]$, where $I_{GaTe}$ is the PL intensity for films with $h_{film}\approx$200 nm.  The absorption coefficient $\alpha_{GaTe}$=5000 cm$^{-1}$ is used according to Ref.\cite{CamasselPRB1979}. A discrepancy by a few orders of magnitude between the experiment and the calculated curve in a wide range of film thicknesses is evident on the graph.

\subsection{Low-temperature cw PL results: GaSe}

Fig.\ref{Fig3}(a) shows typical PL spectra measured for Si$_3$N$_4$-capped films of GaSe of several thicknesses between 8 and 70 nm (the cw laser power of 2 mW is used). The PL signal is observed in a range from 1.95 to 2.05 eV, which is below the emission energy of the free exciton, reported to be around 2.10 eV  for some high purity GaSe samples (see e.g. Ref.\cite{MercierPRB1975}). The detected PL in our samples thus comes from impurity/defect states. The observed localized states may originate from the non-stoichiometric composition of the bulk material.

\begin{figure}[t]
\includegraphics[width=0.95\textwidth]{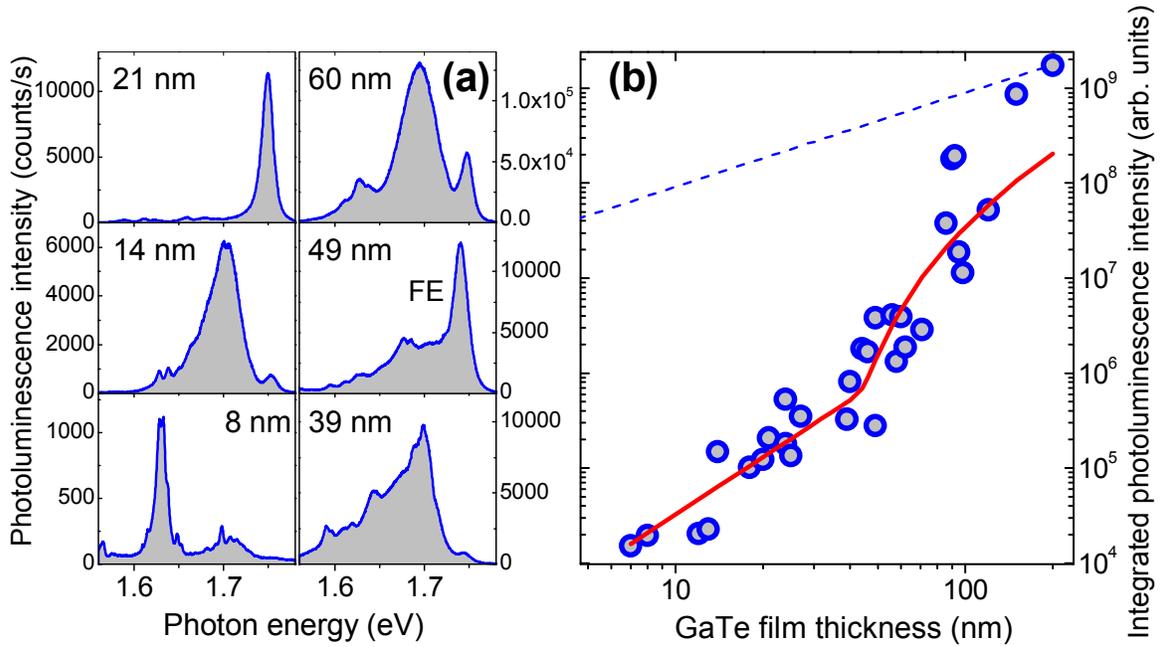}
   \caption{\label{Fig2} Low-temperature cw PL results for GaTe thin films. (a) PL spectra for films of various thicknesses. $h_{film}$ is marked on the plots. On the plot for the film with $h_{film}$=49 nm FE marks the free exciton peak observed at similar energies for all films on the figure except the one with $h_{film}$=8 nm. (b) Symbols show experimentally measured integrated PL (for $T$=10K) for the cw laser excitation power of 2 mW at 532 nm (2.33 eV). Dashed line shows expected variation of PL following the change in the absorption of the thin film assuming constant quantum efficiency. Solid line shows the results of calculations using the model discussed in text.}
\end{figure}

\begin{figure}[t]
\includegraphics[width=0.95\textwidth]{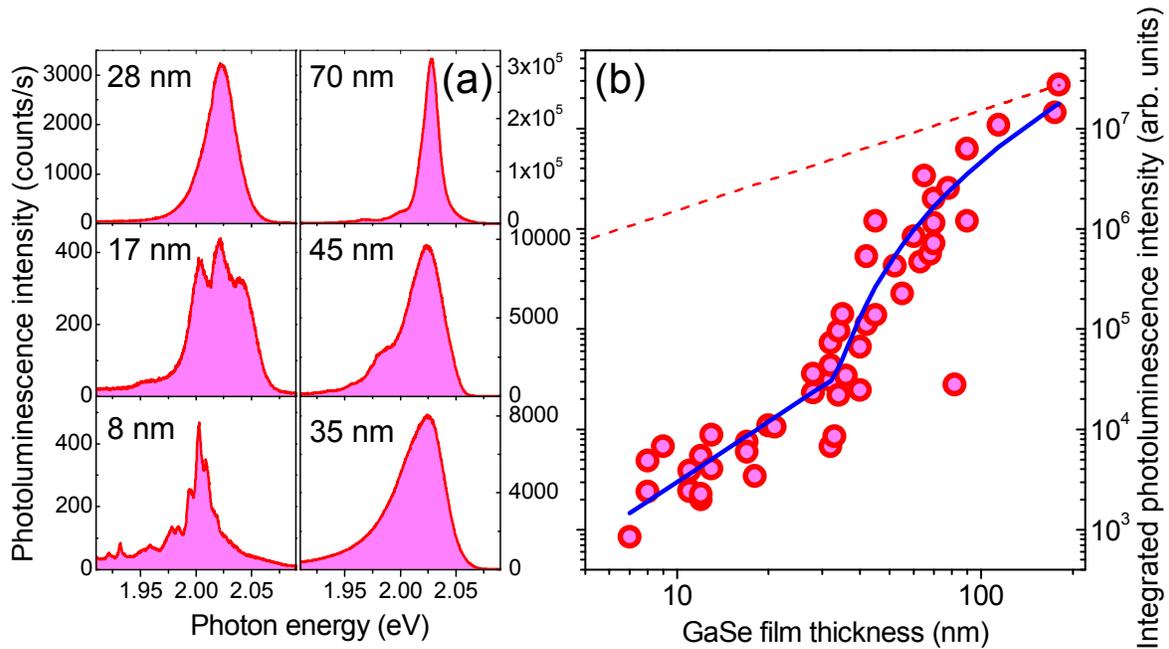}
   \caption{\label{Fig3} Low-temperature cw PL results for GaSe thin films. (a) PL spectra for films of various thicknesses. (b) Symbols show experimentally measured integrated PL (for $T$=10K) for the cw laser excitation power of 2 mW at 532 nm (2.33 eV). Dashed line shows expected variation of PL following the change in the absorption of the thin film assuming constant quantum efficiency. Solid line shows the results of calculations using the model discussed in text.}
\end{figure}

In Fig.\ref{Fig3}(a) it is observed that PL spectra of thin films $<$20 nm usually consist of multiple pronounced lines [a feature similar to GaTe in Fig.\ref{Fig2}(a)], whereas PL spectra tend to exhibit a single pronounced peak for thicker films. PL  linewidths vary between 15 and 60 meV. In some GaSe films, sharp PL lines with linewidths below 5 meV, similar to the sharp features observed in the PL spectrum for the 8 nm film in Fig.\ref{Fig3}(a), also occur in the whole energy range of GaSe PL. From the data measured on more than 50 films, we observe that PL peak energies have a very wide distribution in the range 1.99-2.06 eV. Similarly to GaTe, there is no evidence for size-quantization effects as a function of the film thickness. 

\begin{figure}[t]
\includegraphics[width=0.5\textwidth]{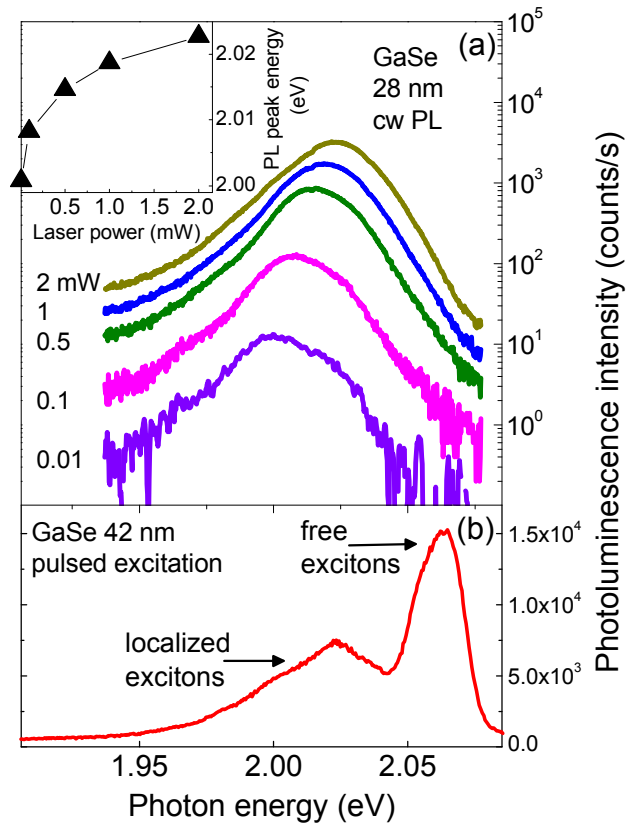}
   \caption{\label{Fig4} (a) Observation of the Stokes shift of localized exciton states in cw PL of a 28 nm thick GaSe films. Excitation with a cw laser at 532 nm (2.33 eV) is used at $T$=10K. (a) Observation of the free exciton feature in PL of a 42 nm thick GaSe films. The time-integrated PL is shown measured for pulsed laser excitation at 420 nm (2.95 eV) with a power of 1 mW. A free exciton peak having a linewidth of $\approx$25 meV is observed at 2.06 eV.}
\end{figure}

Further evidence for exciton localization in thin films is a pronounced Stokes blue-shift observed when the laser excitation density is increased as shown in Fig.\ref{Fig4}. The inset in Fig.\ref{Fig4}(a) shows that the PL peak shifts by $\approx$20 meV as the cw laser power is changed from 0.01 to 2 mW. This is a typical behavior observed in all GaSe films independent on the film thickness: at high power, saturation of some of the PL features is observed accompanied in most cases with a blue-shift of PL of around 10-20 meV. This is a typical behavior observed for localized exciton states in semiconductors, the effect also similar to the state-filling phenomenon in semiconductor quantum dots \cite{RaymondPRB1999}. In some GaSe films, if the optical pumping is further increased, for example, by using pulsed excitation, a relatively broad free exciton feature can be observed, as shown in Fig.\ref{Fig4}(b). This behavior is in agreement with that observed previously in GaTe and GaSe under pulsed excitation, and is related to saturation of the localized states having relatively slow recombination rates \cite{TaylorJofPC1987}. 

Similarly to GaTe films, a significant decrease of PL intensity with the decreasing thickness of the GaSe films is observed [see Fig.\ref{Fig3}(b)] by about 2$\times 10^4$ between 200 and 7 nm. As for the GaTe films in Fig.\ref{Fig2}(b), the strongest PL reduction by 3 orders of magnitude is detected between 200 and 30 nm. For $h_{film}<$30 nm, the PL intensity reduction slows down and is about 30 when $h_{film}$ is varied between 30 and 7 nm. Similar to GaTe in films with $h_{film}<$7 nm PL is completely suppressed. Similarly to Fig.\ref{Fig2}, we show a curve that describes PL reduction due to the reduced absorption only calculated as $I_{PL}=I_{GaSe}[1-exp(-\alpha_{GaSe} h_{film})]$, where $I_{GaSe}$ is the PL intensity for films with $h_{film}$=150 nm and the absorption coefficient $\alpha$=1000 cm$^{-1}$ \cite{CamasselPRB1979}. As for GaTe, a significant discrepancy by a few orders of magnitude between the experimental results and the calculated curve is clear in a wide range of film thicknesses.

\subsection{Time-resolved PL measurements}

\begin{figure}[t]
\includegraphics[width=0.7\textwidth]{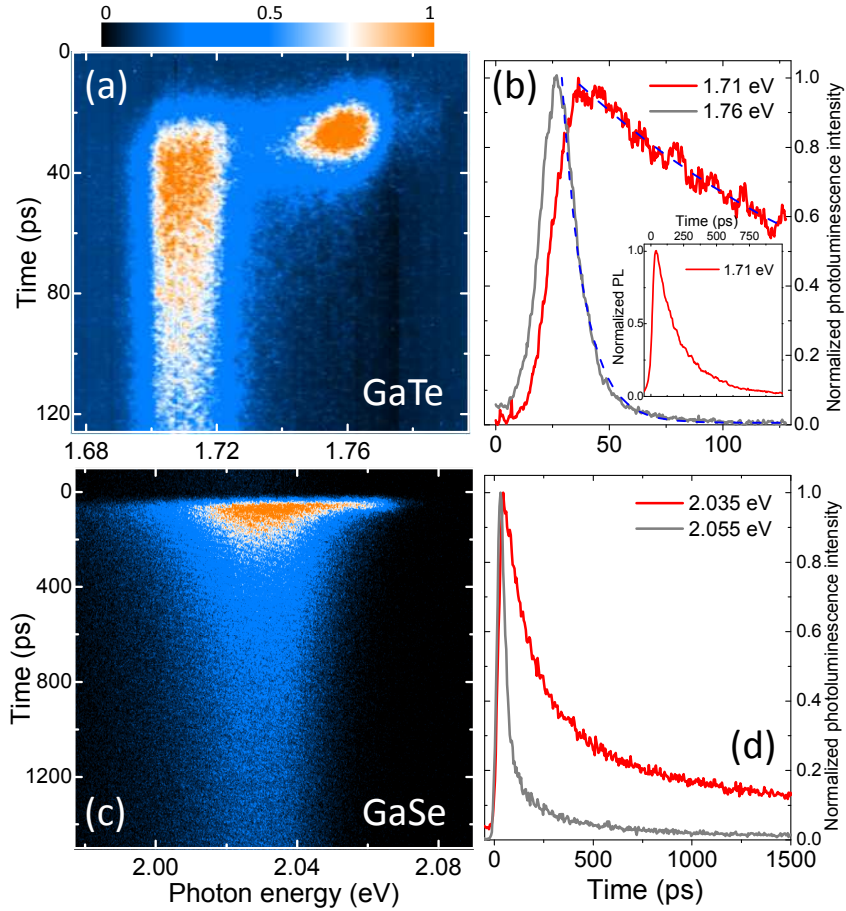}
   \caption{\label{Fig5} Time-resolved PL data for GaTe (a,b) and GaSe (c,d) thin films measured at $T$=10K. Note different time-scales for the GaTe and GaSe data. (a) and (c) show streak-camera scans. (b) and (d) show PL traces measured at 1.76 (gray) and 1.71 eV (red) for GaTe and 2.055 (gray) and 2.035 eV (red) for GaSe. Blue lines in (b) show fitting with a single exponential functions as described in text.}
\end{figure}

In order to shed further light on the results of cw PL and also to provide further experimental foundation for our theoretical model, time-resolved PL experiments have been carried out for a range of films with different thicknesses. Fig.\ref{Fig5} shows typical TRPL data obtained at $T\approx$10K. 

For GaTe, the difference in the origin of the PL features observed in Fig.\ref{Fig2}(a) is further evidenced in TRPL. Fig.\ref{Fig5}(a,b) shows data for a 27 nm thick film exhibiting a behavior typical for films with $h_{film}$ in the range 20 to 200 nm. Fig.\ref{Fig5}(a) presents a streak-camera scan clearly showing two pronounced features at 1.76 eV and 1.71 eV corresponding to the free and localized excitons, respectively. The free exciton peak intensity decays considerably faster than that of the localized states, which does not change significantly on the time-scale of 130 ps shown in the figure. Fig.\ref{Fig5}(b) shows two decay curves measured at 1.76 eV (gray) and 1.71 eV (red). Fitting with single exponential decay functions is shown with blue curves and gives 10 ps for the free exciton and 150 ps for the localized states. The inset shows a PL decay curve measured at 1.71 eV on a larger time-scale, exhibiting an almost complete decay of the signal at 1 ns. We find similar life-times for other films, however no clear dependence on the film thickness is observed: the free exciton PL decay time varies between 5 and 25 ps and that for the localized states between 100 and 200 ps. The lifetimes also weakly depend on the laser excitation power. We also note the rise times of $\approx$15 and 20 ps for the free and localized states, respectively, indicating fast carrier relaxation into the light-emitting states. 

Similar difference between the PL dynamics of the high and low energy part of the spectrum is observed for GaSe thin films in Fig.\ref{Fig5}(c),(d). Here a behavior resembling the Stokes shift shown in Fig.\ref{Fig4} is observed: as the carrier density decreases with time after the laser pulse, the PL intensity maximum progressively moves to lower energy. Fig.\ref{Fig5}(d) details the behavior shown in Fig.\ref{Fig5}(c): two decay curves measured at 2.055 eV (gray) and 2.035 eV (red) are shown. In the center of the PL band at 2.035 eV, the non-exponential decay occurs with a characteristic time of 400 ps, which also shows a slow-decaying component. At around 2.055 eV, the initial PL time-dependence  can be well fitted with a single-exponential decay with a lifetime of $\approx$40 ps, dominated most likely by carrier relaxation to lower energy. The complex behavior in GaSe films occurs due to the partial saturation of the states at short times after the excitation pulse and fast relaxation to lower energy. We find similar behavior for films with other thicknesses. Similarly to GaTe no clear dependence on the film thickness is observed and the rise times of $\approx$30 ps are found for the localized states.

\subsection{Modeling}

\begin{figure}[t]
\includegraphics[width=0.8\textwidth]{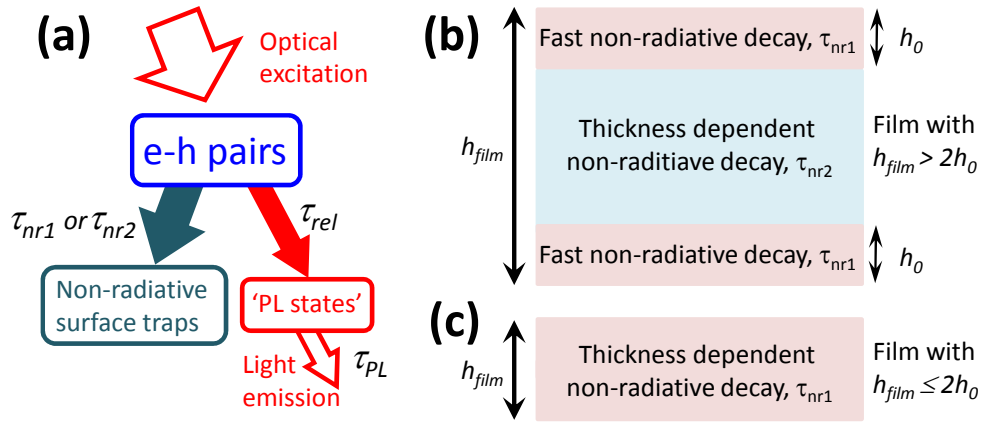}
   \caption{\label{Fig6} Diagrams illustrating the theoretical model for the dependence of the cw PL intensity on the film thickness. (a) Processes included in the model (see text and supplementary information for more details): optical excitation of e-h pairs; relaxation into the non-radiative traps with a time $\tau_{nr1}$ or $\tau_{nr2}$ depending on the thickness of the film; relaxation with a time $\tau_{rel}$ into the light-emitting states denoted as 'PL states'; PL emission from these states with a time $\tau_{PL}$. (b) and (c) shows two types of thin films with thicknesses above and below the critical thickness of $2h_0$. In (a) we show a relatively thick film with a thickness of $h_{film}>2h_0$. This film has three regions: two regions of thickness $h_0$ near the film surfaces where fast carrier relaxation to surface traps occurs (shown in pink), and a 'normal' region of thickness $h_{film}-2h_0$ in the middle of the film (blue) where non-radiative processes are weaker and occur through carrier escape into the surface regions. In (b) we show a thin film with $h_{film}\leq 2h_0$, where a region of one type only exists, where fast non-radiative carrier decay occurs.}
\end{figure}

In order to describe the observed trend of PL intensity as a function of film thickness $h_{film}$ in GaSe and GaTe thin films we have developed a rate equation model presented in detail in the Supplementary Information (SI)\cite{SI}. As shown in Fig.\ref{Fig6}(a), in the model we consider the following processes leading to light emission: optical excitation of e-h pairs (uniformly across the full thickness of the film); relaxation into the non-radiative traps with a time $\tau_{nr1}$ or $\tau_{nr2}$ depending on the thickness of the film, as explained below; relaxation with a time $\tau_{rel}$ into the light-emitting states denoted in Fig.\ref{Fig6}(a) as 'PL states'; PL emission from these states with a time $\tau_{PL}$. The absence of a clear dependence of the PL decay times on the film thickness allows us to use a thickness-independent constant $\tau_{PL}$ and neglect any other decay mechanisms for the e-h population of the light-emitting states. The physical origin of this may be in a relatively strong localization of e-h pairs in the light-emitting 'PL states', which leads to suppression of non-radiative processes. We also assume that the mechanism of relaxation into the 'PL states' is not sensitive to the film thickness and can be described with a thickness independent time $\tau_{rel}$.

As shown in Fig.\ref{Fig6}(b) and (c) we also assume that the film is divided in three regions: (1) two regions of thickness $h_0$ near the film surfaces where fast carrier relaxation to surface traps occurs leading to non-radiative carrier escape; (2) a 'normal' region of thickness $h_{film}-2h_0$ in the film's central part not containing the traps where the photo-excited carriers can escape into regions (1), where they undergo non-radiative decay. Here the $h_0$ value may be associated with a depletion depth or an average surface trap radius\cite{CalarcoNanoLett2005}. As shown in Fig.\ref{Fig6} there are two possibilities: in Fig.\ref{Fig6}(b), where $h_{film}>2h_0$ and both regions of type (1) and (2) exist; and in  Fig.\ref{Fig6}(c), for films with $h_{film} \leq 2h_0$, where region (2) is not present. The model assumes that light absorption and PL occurs in both types of regions. In both regions, e-h pairs relax with the time $\tau_{rel}$ into the light-emitting states.  

In region (1), in first approximation the average time it takes for the carrier/e-h pair to escape non-radiatively is proportional to half the thickness of region (1) (can be understood as the average time for the carrier to reach the surface or as the overlap of the wavefunction of the carrier and the surface trap). In region (2), the non-radiative escape time reflects the average time it takes for a carrier or an e-h pair to reach any of the regions (1).  The underlying mechanism for this process may be depletion and band-bending expected at the film surface leading to charge separation and non-radiative decay \cite{CalarcoNanoLett2005}. We assume that once the carrier or e-h pair has reached region (1) it escapes non-radiatively. In first approximation, the average time it takes a carrier/e-h pair to reach region (1) is proportional to half the thickness of region (2). Thus we introduce non-radiative decay times in region (1) as $\tau_{nr1}=(h_0/2)/u_1$ for $h_{film} > 2h_0$, and $\tau_{nr1}=(h_{film}/4)/u_1$ for $h_{film} \leq 2h_0$. In region (2) it is $\tau_{nr2}=(h_{film}/2-h_0)/u_2$. Here $u_1$ and $u_2$ have dimensions of m/s. In the case of region 2 where effectively we assume ballistic exciton (or electron/hole) transport preceding the non-radiative escape, $u_2$ can be interpreted as the average carrier velocity.

\subsection{Discussion}

As detailed in Supplementary Information\cite{SI} and observed in Fig.\ref{Fig2}(b) and Fig.\ref{Fig3}(b), we find that the proposed model provides a reasonable description of our data using four parameters (which are not completely independent as we find): the 'critical' thickness $h_0$, a parameter describing the amount of light absorbed by the film, and products $\tau_{rel}u_1$ and $\tau_{rel}u_2$. In particular the fitting functions that we produce capture the change in the 'slope' of the data observed at around 30 nm for GaSe and 40 nm for GaTe, which we interpret as the thickness of the film where $h_{film}\approx2h_0$, i.e. the thickness of region (2) turns to zero, and non-radiative carrier escape changes its character. 

However, we find that the accuracy of the fitting is not sufficient to extend our analysis beyond determination of the order of magnitude of the products $\tau_{rel}u_1$ and $\tau_{rel}u_2$. We find that $\tau_{rel}u_1$ and $\tau_{rel}u_2$ are of the order of 1000 nm and 100 nm, respectively, for both GaSe and GaTe. The solid line shown in Fig.\ref{Fig2}(b) for GaTe films is obtained for $\tau_{rel}u_1 = 3800$ nm and  $\tau_{rel}u_2= 200$ nm, whereas the fitting in Fig.\ref{Fig3}(b) for GaSe films is done for $\tau_{rel}u_1 = 2900$ nm and  $\tau_{rel}u_2= 150$ nm. We note that the description of GaSe PL is more satisfactory, possibly because in GaTe there is a contribution from free exciton PL, so additional non-radiative escape channels and relaxation processes need to be taken into account. For thicker GaTe films, much stronger PL than predicted by the model is observed, which probably signifies suppression of additional non-radiative escape that free excitons experience in relatively thin films. It is also notable that in GaTe the PL lifetime for localized states is shorter than in GaSe, which may be due to non-radiative escape. Such processes are not included in the model. Another reason could be deviation from the effectively 'ballistic' transport that we assume leads to the carrier escape into regions (1), and its replacement for the larger thicknesses with a slower 'diffusion' process leading to slower non-radiative escape.  

Assuming that the measured PL rise-times of $\approx20$ ps are close to $\tau_{rel}$, we can estimate characteristic non-radiative times $\tau_{nr1}$ and $\tau_{nr2}$ for several limiting cases. For example for GaSe we obtain the following values using $\tau_{rel}u_1$=1000 nm and $\tau_{rel}u_2$=100 nm: for $h_{film}$=10 nm $\tau_{nr1}$=0.05 ps, for $h_{film}$=30 nm $\tau_{nr1}$=0.15 ps, for $h_{film}$=100 nm $\tau_{nr2}$=7 ps. For both $h_{film}$ of 10 and 30 nm, the non-radiative decay occurs on a sub-picosecond time-scale. Here, the non-radiative escape is by a factor of the order of 100 faster than relaxation into the light emitting states, which is consistent with a low quantum yield of 10$^{-3}$ reported previously for thin films of MoS$_2$\cite{MakPRL2010}. For the middle region of a 100 nm film, the characteristic non-radiative escape time is comparable with the relaxation time into the states giving rise to PL, thus a much higher quantum yield can be expected for films of this thickness.   

We also note that in the case of region (2), $u_2$ should be of the order of a typical thermal velocity of an exciton, $v_{th}$. By using $T$=10 K and the exciton mass of 0.1 (or 0.2)\cite{WatanabePRB2003} of the free electron mass we obtain $\tau_{rel} v_{th}\approx$ 1200 (or 900) nm for $\tau_{rel}\approx$20 ps, very similar to the order of magnitude predicted by the model for the product $\tau_{rel}u_2$.

\subsection{Conclusions}

We have investigated thin films of GaTe and GaSe prepared by mechanical exfoliation from bulk crystals, deposited on SiO$_2$ substrates and capped with a thin layer of Si$_3$N$_4$. The study of optical properties of the thin films has been conducted by means of low-temperature cw and time-resolved micro-PL techniques. The most pronounced property observed is a significant reduction of cw PL intensity by up to 10$^5$ for thin films of thicknesses about 10 nm compared with the films of 150-200 nm. No measurable PL was observed in films thinner than 7 nm. Except for PL decrease, no other clear trends, including size-quantization effects and PL life-time modifications as a function of the film thickness were observed in cw and time-resolved PL. We argue that the reduction of quantum yield occurs due to the non-radiative processes associated with the surface states. The theoretical model that we develop differentiates between the fast non-radiative carrier escape to surface traps in the thin 15-20 nm layers near the film surface, and slower decay in the middle region of the film, where the carriers first move to the surface layers and then decay. The model accounts for the change in the character of the PL decay for thin films with thicknesses less than 30-40 nm, where we expect only the fast direct relaxation to surface traps. 

We do not refer in our model and interpretation to the direct-to-indirect bandgap transition found for thin InSe films in Ref.\cite{MuddAdvMat2013}. The observed strong PL decrease is not accompanied with any significant variation of the PL lifetime, which would most likely accompany such a change in the band structure. This implies that the PL reduction in our case is not associated with this type of transition, and instead the reduction in the radiative recombination contribution is likely due to an increase in the non-radiative decay rate in thin films. 

The strong non-radiative decay processes occur in the studied films despite complete encapsulation in Si$_3$N$_4$ and SiO$_2$ providing partial protection of the surface from chemical interactions with the ambient atmosphere. This emphasizes the need for development of novel surface passivation for III-VI films, possibly involving oxygen free substrates and encapsulation in additional layered materials such as boron nitride. Using this fabrication methods, the large family of III-VI materials may show weaker dependence of quantum yield on film thickness, which will allow them to play important role as building blocks in van der Waals heterostructures. It is likely that in the near future fabrication of such heterostructures will be carried out in oxygen- and water-free atmospheres and will include carefully designed passivating layers.  

\begin{acknowledgement}

We thank the financial support of EPSRC grant EP/M012700/1, Graphene Flagship project 604391, FP7 ITN S3NANO, EPSRC Programme Grant EP/J007544/1, SEP-Mexico and CONACYT. \\ 

\end{acknowledgement}


\providecommand{\latin}[1]{#1}
\providecommand*\mcitethebibliography{\thebibliography}
\csname @ifundefined\endcsname{endmcitethebibliography}
  {\let\endmcitethebibliography\endthebibliography}{}

\pagebreak

\pagebreak[4]

\pagebreak

\pagebreak

\pagebreak

\clearpage
\renewcommand{\thefigure}{S\arabic{figure}}
\setcounter{figure}{0}
\renewcommand{\thetable}{S\arabic{table}}
\setcounter{table}{0}
\renewcommand{\citenumfont}[1]{S#1}
\makeatletter
\renewcommand{\@biblabel}[1]{S#1.}
\makeatother
\pagebreak \pagenumbering{arabic}

\clearpage

\begin{center}
\section*{Optical properties of two-dimensional gallium chalcogenide films: Supplementary Information}
\end{center}

\hspace{2cm}

\begin{center}
\textbf{Abstract}
\end{center}

Here we present additional details on the rate-equation model describing the thickness dependence of photoluminescence intensity in GaSe and GaTe thin films.


\section{Modeling}

In order to describe the observed trend of photoluminescence (PL) intensity as a function of film thickness $h_{film}$ in GaSe and GaTe thin films we have developed a simplified rate equation model described below. It describes the behavior of photo-excited carriers in terms of populations of e-h pairs, i.e. electrons and holes are not treated separately for simplicity. 

\begin{figure}[t]
\includegraphics[width=0.8\textwidth]{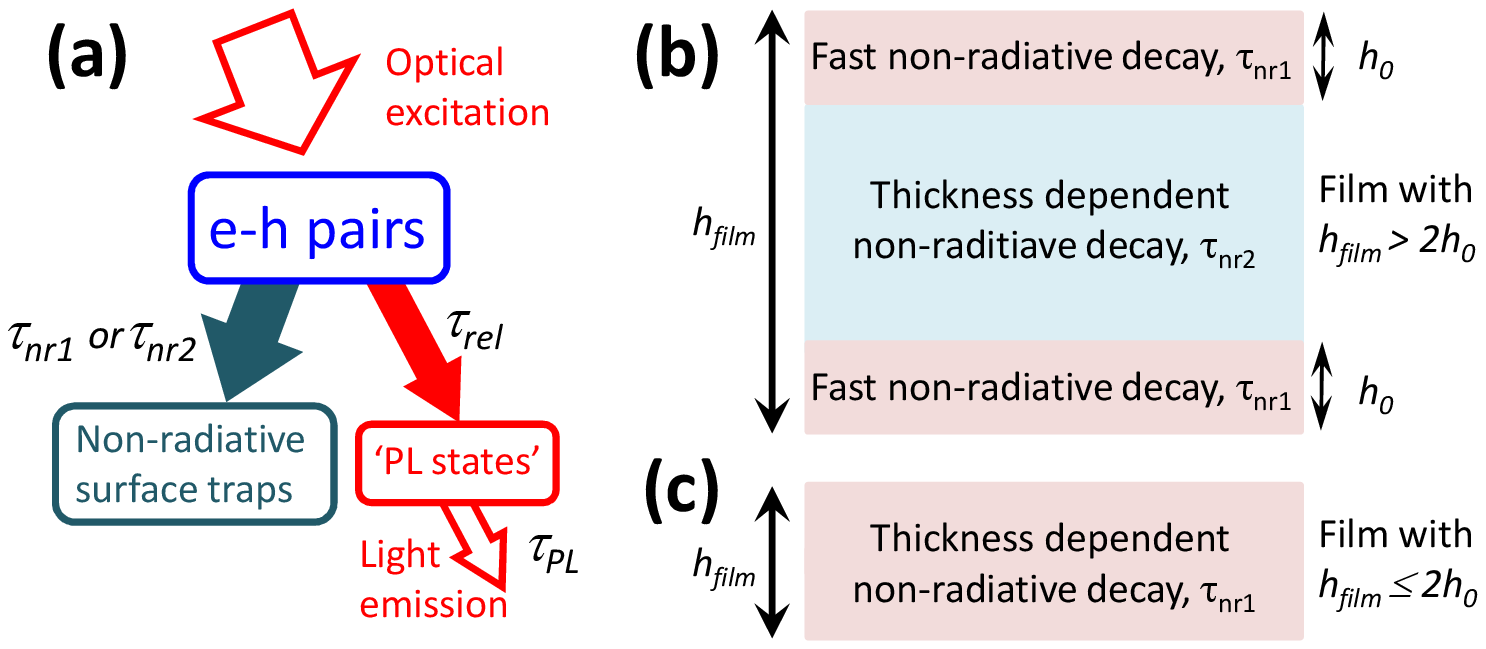}
   \caption{\label{Fig1SI} Diagrams illustrating the theoretical model for the dependence of the cw PL intensity on the film thickness. (a) Processes included in the model (see text and supplementary information for more details): optical excitation of e-h pairs; relaxation into the non-radiative traps with a time $\tau_{nr1}$ or $\tau_{nr2}$ depending on the thickness of the film; relaxation with a time $\tau_{rel}$ into the light-emitting states denoted as 'PL states'; PL emission from these states with a time $\tau_{PL}$. (b) and (c) shows two types of thin films with thicknesses above and below the critical thickness of $2h_0$. In (a) we show a relatively thick film with a thickness of $h_{film}>2h_0$. This film has three regions: two regions of thickness $h_0$ near the film surfaces where fast carrier relaxation to surface traps occurs (shown in pink), and a 'normal' region of thickness $h_{film}-2h_0$ in the middle of the film (blue) where non-radiative processes are weaker and occur through carrier escape into the surface regions. In (b) we show a thin film with $h_{film}\leq 2h_0$, where a region of one type only exists, where fast non-radiative carrier decay occurs.}
\end{figure}

As shown in Fig.\ref{Fig1SI}(a), in the model we consider the following processes leading to light emission: optical excitation of e-h pairs (uniformly across the full thickness of the film); relaxation into the non-radiative traps with a time $\tau_{nr1}$ or $\tau_{nr2}$ depending on the thickness of the film, as explained below; relaxation with a time $\tau_{rel}$ into the light-emitting states denoted in Fig.\ref{Fig1SI}(a) as 'PL states'; PL emission from these states with a time $\tau_{PL}$. The absence of a clear dependence of the PL decay times on the film thickness allows us to use a thickness-independent constant $\tau_{PL}$ and neglect any other decay mechanisms for the e-h population of the light-emitting states. We also assume that the mechanism of relaxation into the 'PL states' is not sensitive to the film thickness and can be described with a thickness independent time $\tau_{rel}$.

As shown in Fig.\ref{Fig1SI}(b) and (c) we assume that the film is divided into three regions: (1) two regions of thickness $h_0$ near the film surfaces where fast carrier relaxation to surface traps occurs leading to non-radiative carrier escape; (2) a 'normal' region of thickness $h_{film}-2h_0$ in the central part of the film where carriers do not decay non-radiatively, but can escape into the regions (1), where they undergo non-radiative decay. Here the $h_0$ value may be associated with a depletion depth or an average surface trap radius\cite{CalarcoNanoLett2005}. It is assumed to be independent of the film thickness and may be different for films made of different materials. 

In region (1), in first approximation the average time it takes for the carrier/e-h pair to escape non-radiatively is proportional to half the thickness of region (1) (can be understood as the average time for the carrier to reach the surface or as the overlap of the wavefunction of the carrier and the trap). This half-thickness is given by $h_0/2$ for the two regions adjacent to the surface for film thicknesses $h_{film} > 2h_0$, and is $h_{film}/4$ for $h_{film} \leq 2h_0$, when the two surface regions merge. Thus we introduce a non-raditive decay time in the regions of type (1) as $\tau_{nr1}=(h_0/2)/u_1$ for $h_{film} > 2h_0$, and $\tau_{nr1}=(h_{film}/4)/u_1$ for $h_{film} \leq 2h_0$. Here $u_1$ is a constant with the dimensions of meter per second.

In region (2), the non-radiative escape time reflects the average time it takes for a carrier or an e-h pair to reach any of the regions (1).  The underlying mechanism for this process may be depletion and band-bending expected at the film surface leading to charge separation and non-radiative decay \cite{CalarcoNanoLett2005}. We assume that once the carrier or e-h pair has reached region (1) it escapes non-radiatively. In first approximation, the average time it takes a carrier/e-h pair to reach region (1) is proportional to half the thickness of region (2), $h_{film}/2-h_0$. Thus we introduce a non-raditive decay time in region (2) as $\tau_{nr2}=(h_{film}/2-h_0)/u_2$. Here $u_2$ is a constant with the dimensions of meter per second, effectively corresponding to the average carrier velocity in region (2). Notably, in films with $h_{film} \leq 2h_0$, region (2) does not exist, and the only non-radiative escape that we consider is due to the escape to traps as described above for region (1).

\begin{figure}[t]
\includegraphics[width=0.5\textwidth]{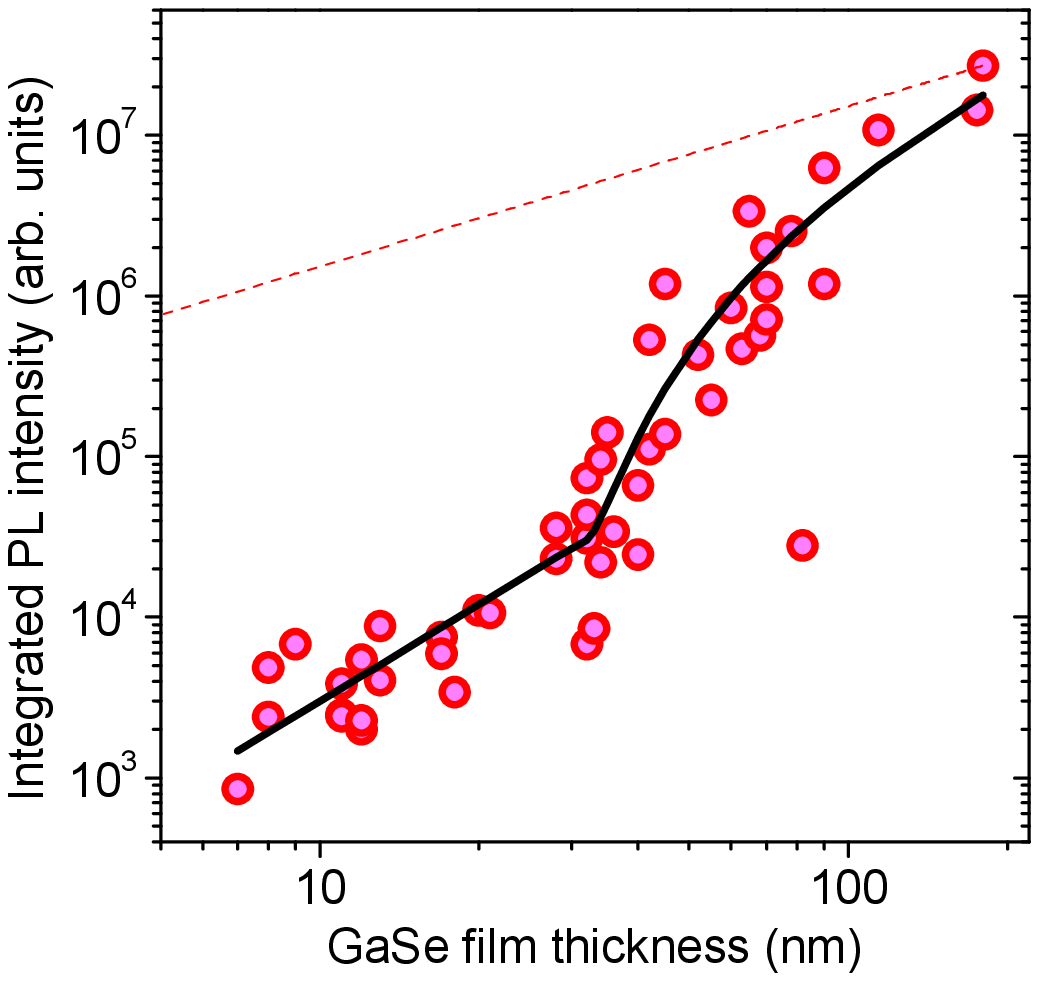}
   \caption{\label{Fig2SI} Low-temperature cw PL results for GaSe thin films. Symbols show experimentally measured integrated PL (for $T$=10K) for the laser excitation power of 2 mW (laser excitation at 532 nm is used). Dashed line shows expected variation of PL following the change in the absorption of the thin film. Solid line shows the results of calculations using the model discussed in text.}
\end{figure}

In equations below, we denote e-h pair populations in regions (1) and (2) as $N_1$ and $N_2$, respectively, and the population of the 'PL states' as $N_{loc}$. Appropriately, $N_2=0$ for $h_{film} \leq 2h_0$. In both regions, e-h pairs relax with the time $\tau_{rel}$ into the 'PL states' giving rise to radiative recombination. Non-radiative escape from the 'PL states' is neglected. This assumption is based on the fact that in most cases PL is observed from the tightly-bound localized states. $N_{loc}$ states decay radiatively with a characteristic time $\tau_{PL}$, so that the PL intensity $I_{PL}=N_{loc}/\tau_{PL}$. Saturation of 'PL states' is neglected as the data reported in Figs.\ref{Fig2SI},\ref{Fig3SI} is measured for the laser powers where saturation effects are unimportant. The light absorption coefficient, $\alpha$ is assumed to be the same in both types of regions (1) and (2).  For the small thicknesses of the films that we consider, the e-h pair generation rate is equal to $P\alpha$  multiplied by the thickness of the regions (1) or (2) as given below. Here $P$ is the e-h pair generation rate due to the laser excitation rescaled to account for the PL detection efficiency.  

The following rate equations can thus be introduced for the case $h_{film} > 2h_0$:  

$$
\begin{array}{lcl}
{{dN_1}/{dt}}  &=  & 2h_0P\alpha-N_1/t_{rel}-N_1/\tau_{nr1}\\
{{dN_2}/{dt}}  &=  &(h_{film} - 2h_0)P\alpha-N_2/t_{rel}-N_2/\tau_{nr2}\\
{{dN_{loc}}/{dt}}  &=  &N_1/\tau_{rel}+ N_2/\tau_{rel}- N_{loc}/\tau_{PL}
\end{array}
$$

For the case $h_{film} \leq 2h_0$ we will have:  

$$
\begin{array}{lcl}
{{dN_1}/{dt}}  &=  &P h_{film}\alpha-N_1/t_{rel}-N_1/\tau_{nr1}\\
N_2  &=  &0\\
{{dN_{loc}}/{dt}}  &=  &N_1/\tau_{rel} - N_{loc}/\tau_{PL}
\end{array}
$$

In the steady-state case we obtain the following expression for PL intensity for $h_{film} > 2h_0$:

\begin{equation}
I_{PL}=2h_0P\alpha/(1+\tau_{rel}/\tau_{nr1}) + (h_{film} - 2h_0)P\alpha/(1+\tau_{rel}/\tau_{nr2}),
\end{equation}

where $\tau_{nr1}=\tau_1h_0/2$ and $\tau_{nr2}=\tau_2(h_{film}/2-h_0)$, and for $h_{film} \leq 2h_0$

\begin{equation}
I_{PL}=h_{film}P\alpha/(1+\tau_{rel}/\tau_{nr1}),
\end{equation}

where $\tau_{nr1}=(h_{film}/4)/u_1$. 

By substituting these expressions for the non-radiative decay times directly we obtain for PL intensity for $h_{film} > 2h_0$:

\begin{equation}
\label{I_PL_thick}
I_{PL}=2h_0^2P\alpha/(h_0+2\tau_{rel}u_{1}) + (h_{film} - 2h_0)^2P\alpha/(h_{film} - 2h_0+2\tau_{rel}u_{2}).
\end{equation}

For $h_{film} \leq 2h_0$ we get: 

\begin{equation}
\label{I_PL_thin}
I_{PL}=h_{film}^2P\alpha/(h_{film}+4\tau_{rel}u_{1}),
\end{equation}

\section{Results of the modeling}

\subsection{Results for GaSe thin films}

According to Eqs. \ref{I_PL_thick}, \ref{I_PL_thin}, we have four parameters, with which we describe the observed dependence of $I_{PL}$ on the film thickness: $h_0$, the products $P\alpha$, $\tau_{rel}u_1$ and $\tau_{rel}u_2$. 

Further analysis of Eqs. \ref{I_PL_thick},\ref{I_PL_thin} shows that these parameters are not completely independent. This becomes obvious if we note that for small $h_{film}$ where we expect fast non-radiative process (i.e. large $\tau_{rel}u_{1}$) we get  $h_{film} \ll 4\tau_{rel}u_{1}$, and Eq.\ref{I_PL_thin} reduces to a simple parabolic dependence: $I_{PL}\approx h_{film}^2P\alpha/(4\tau_{rel}u_{1})$. This function is independent of individual parameters $P\alpha$ and $\tau_{rel}u_1$, and depends on their ratio only. 

For large $h_{film}$ where we expect slower non-radiative process we get  $(h_{film}- 2h_0)$ of the order of $2\tau_{rel}u_{2}$, and a simple approximation can be used for Eq. \ref{I_PL_thick} giving $I_{PL}\approx (h_{film} - 2h_0)P\alpha$. From here we find that $P\alpha$ should be of the order of 10$^5$ nm$^{-1}$s$^{-1}$. 

The two functions described by Eqs. \ref{I_PL_thick}, \ref{I_PL_thin} and especially their asymptotics have significantly different form, effectively with different 'slopes' as seen e.g. in Fig.\ref{Fig2SI} for the case of GaSe. For GaSe data in Fig.\ref{Fig2SI}, it is relatively easy to define the position of a characteristic kink around $h_{film}=$30 nm, where the behavior changes. We use this observation to fix the value of $h_0$ to 15 nm.

In the next step, we have explored a range of fitting functions providing a reasonable agreement with the experiment. We conclude that our model can reliably predict the order of magnitude of the fitting parameters, but not the actual value. So we find that $\tau_{rel}u_1$ is of the order of a 1000 nm and $\tau_{rel}u_2$ is of the order of a 100 nm.
 
As an illustration for the GaSe data, in Fig.\ref{Fig2SI} we show a fitting with $P\alpha=3.5 \times 10^5$nm$^{-1}$s$^{-1}$, $\tau_{rel}u_1 = 2900$ nm and  $\tau_{rel}u_2= 150$ nm. The figure also shows the variation of PL intensity as would be expected from the change in absorption only (dashed line). This would correspond to a thickness independent non-radiative processes occurring with the same rate in the whole volume of the film. This curve is described by the expression $I_{PL}=I_{GaSe}[1-exp(-\alpha_{GaSe} h_{film})]$, where $I_{GaSe}$ is the PL intensity for films with $h_{film}\approx$200 nm and the absorption coefficient $\alpha$=1000 cm$^{-1}$ \cite{CamasselPRB1979}.

\begin{figure}[t]
\includegraphics[width=0.5\textwidth]{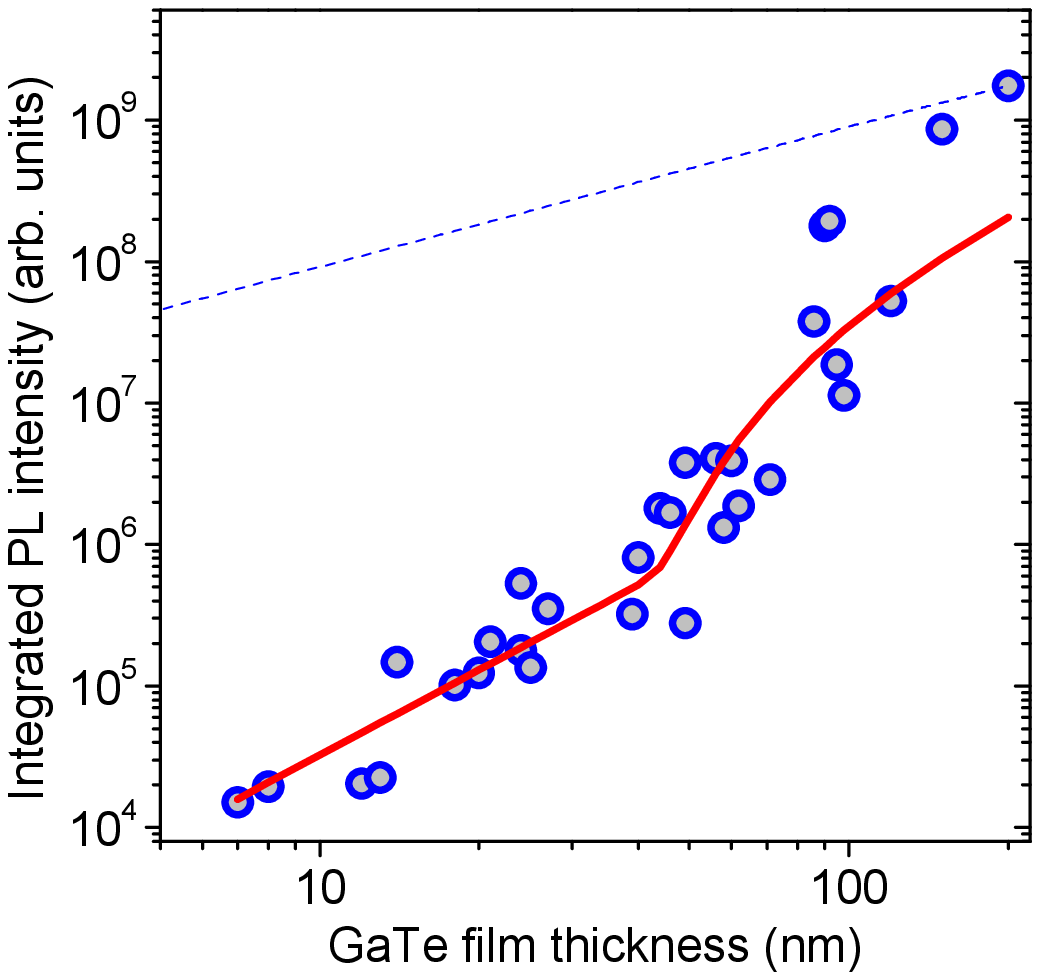}
   \caption{\label{Fig3SI} Low-temperature cw PL results for GaTe thin films. Symbols show experimentally measured integrated PL (for $T$=10K) for the laser excitation power of 2 mW (laser excitation at 532 nm is used). Dashed line shows expected variation of PL following the change in the absorption of the thin film. Solid line shows the results of calculations using the model discussed in text.}
\end{figure}

\subsection{Results for GaTe thin films}

We used the same fitting procedure for GaTe with the results shown in Fig.\ref{Fig3SI}. Although the 'kink' in the thickness-dependence of $I_{PL}$ for GaTe is not as pronounced as for GaSe, it is still observable at a slightly higher value of $2h_0$=40 nm. The model describes adequately the data for $h_{film}$ up to around 90 nm, but fails to reproduce the very steep enhancement of PL for thicker films. This probably occurs due to the contribution of the free excitons to the PL signal. It is probable that there is an additional strong non-radiative escape mechanism for free excitons in thin films, which is overcome in thicker films with the result that the PL intensity for thick films grows faster than the model predicts. Another reason could be deviation from the effectively 'ballistic' transport that we assume leads to the carrier escape into regions (1), and its replacement for the larger thicknesses with a slower 'diffusion' process.

In a similar way to GaSe, we confine ourselves to determination of the order of magnitude only for the main parameters, which are similar to GaSe: we find that $\tau_{rel}u_1$ is of the order of a 1000 nm and $\tau_{rel}u_2$ is of the order of a 100 nm. As an illustration for the GaTe data, in Fig.\ref{Fig3SI} we show a fitting with $P\alpha=5.0 \times 10^6$ nm$^{-1}$s$^{-1}$, $\tau_{rel}/\tau_1 = 3800$ nm and $\tau_{rel}/\tau_2= 200$ nm. The figure also shows the variation of PL intensity as would be expected from the change in absorption only (dashed line). This curve is described by the expression $I_{PL}=I_{GaTe} [1-exp(-\alpha_{GaTe} h_{film})]$, where $I_{GaTe}$ is the PL intensity for films with $h_{film}\approx$200 nm. The absorption coefficient $\alpha_{GaTe}$=5000 cm$^{-1}$ is used according to Ref. \cite{CamasselPRB1979}.


\providecommand{\latin}[1]{#1}
\providecommand*\mcitethebibliography{\thebibliography}
\csname @ifundefined\endcsname{endmcitethebibliography}
  {\let\endmcitethebibliography\endthebibliography}{}

\end{document}